\documentclass[12pt]{iopart}

\usepackage{graphicx}
\begin{document}

\title[A Simple Setup for Cosmic Muon Lifetime Measurements ]
{A Simple Setup for Cosmic
 Muon Lifetime Measurements}

\author{
D. Bosnar$^1$, Z. Mati\'{c} $^2$,  I. Fri\v{s}\v{c}i\'{c}$^{1,*}$, P. \v{Z}ugec$^1$, and H. Jan\v{c}i$^3$}

\address{$^1$ Department of Physics, Faculty of Science, University of Zagreb,
 Croatia \\
$^2$ Metallic Nova d.o.o., Rijeka \\
  $^3$ Osnovna \v{s}kola Grgura Karlov\v{c}ana, Djurdjevac  }
\ead{bosnar@phy.hr}
\begin{abstract}
Elementary particle physics is a fascinating field of modern physics investigating the basic 
constituents of matter and their interactions. In the experiments large accelerators and
very sophisticated  detector systems are usually used. However, it is desirable to have simple experiments
for undergraduate and even secondary school courses which can demonstrate complex investigations in
this field. We have constructed a simple setup for the measurement of the lifetime of cosmic
muons based on a single scintillation detector. Expensive and complicated professional
particle physics equipment for the signal processing, which mainly prevented the realization of this
experiment by non-experts, is replaced by simple, inexpensive and commercially available
electronic components. With our setup
we register time stamps of the events detected in the scintillation detector and from this
data we determine the muon lifetime. Also, Python software package has been developed 
for data analysis and presentation of the results.

\end{abstract}
\footnote{$^*$ Present address:MIT-LNS, Cambridge MA, 02139, USA} 

\maketitle

\section{Introduction}
 Elementary particle physics is investigating properties of
 the basic building blocks of matter and their interactions. 
 According to the current knowledge described by the standard
 model of particle physics, basic constituents of matter are 
 fermions - quarks and leptons - and the interactions between them are
 mediated by bosons, so-called gauge bosons, see e.g. \cite{standard}. 
 Typical experiments in the elementary particle
 physics include investigations of the products of particles collisions or 
 decays of unstable particles.
 Usually, big accelerator
 facilities, where Large Hadron Collider at CERN is an extreme 
 example \cite{cern}, and very complex detector systems are used.
 
 However, it is desirable to have adequate laboratory experiments 
 for undergraduate and even advanced secondary school courses in order
 to demonstrate experimental techniques in this exciting field. The 
 measurement of cosmic muons lifetime has for a long time been considered
 as one of such experiments, see e.g. \cite{mel}.
 In this experiment, cosmic rays provide a free and everywhere available
 source of muons. The muon is  one of the unstable leptons and its lifetime, 
 one of the basic particle properties, is measured. There have been various
 realizations of that experiment, mostly based on scintillation
 detectors  \cite{hall,owens,lewis,ward,riggi}.
 However, all these setups use expensive and
 pretty demanding ``professional'' particle physics electronics for signal
 processing and storage which requires special expertise, and
 also with correspondingly high costs. 
 The foregoing has prevented the realization of this measurement by  non-experts
 and much broader employment of this experiment, even in the secondary schools.
 Nevertheless, some compact setups based on low-cost electronics have also started
 to appear \cite{coan}. 
 
 Exploiting the advances in electronics, we have constructed a simple setup for
 the measurement of the lifetime of cosmic muons.
 It consists of a single scintillation detector for the detection of both muons
 and electrons (or positrons) from
 the decays of the muons which stop in the scintillator.
 The scintillation detector
 is coupled to the electronics that registers time
 stamps of signals in the detector 
 above a certain threshold.
 The time resolution is 12.5 ns and these data are stored either on
 a memory card or on a computer hard disk.  The time
 differences between the two subsequently registered events in the detector
 can be determined from these time stamps.
 Suitable cuts can be applied  on these time differences
 in order to extract time distribution of events that corresponds to muon decays,
 and subsequently the mean lifetime of muons, loosely also called muon lifetime,
 can be determined.
 
 Using Python software we have also developed a graphical user interface for the
 presentation of data, data analysis and presentation of the results.
 Depending on the level of students, they can either develop their own analysis software
 or use the existing analysis tools.
 
 There are various educational topics that can be 
 discussed in the connection with this experiment, such as relativistic time dilation, muon
 lifetime in matter, Fermi weak interaction coupling, parity violation in weak
 interactions and existence of two different types of neutrino \cite{riggi}.
 Since our setup registers time stamps of the events detected in the scintillation
 detector, it is also possible to study the statistics of random events,
 and this topic will be treated in a separate article.
 
 In the second section we briefly review the properties of muons and cosmic muons.
 In the third section we 
 describe 
 principles of our measurement of the muon lifetime. In the fourth section a description of experimental setup
 and measurements is given, and in the fifth section data analysis and results are presented.

\section{Muons and Cosmic Rays Muons}
Muon is an unstable lepton, approximately 206 times heavier than the more 
familiar electron, with half integer spin 1/2. 
Its charge can be either positive or negative. Negative muon decays, by the weak interaction,
almost exclusively into an electron (or positron in the case of positive muon) 
and two corresponding neutrinos/antineutrinos \cite{pdg}:

\begin{eqnarray}
\mu^- \rightarrow e^- + \bar{\nu_e} + \nu_{\mu}   \nonumber \\
\mu^+ \rightarrow e^+ + \nu_e + \bar{\nu_{\mu}}   \nonumber 
\end{eqnarray}

Muons were discovered as new particles in cosmic rays
in the thirties of the last century \cite{and37}, and their properties  have been subsequently
investigated in cosmic rays and accelerator experiments. 
Although muon has been known for almost eighty years, the precise measurement of 
its lifetime is still a topic of scientific research, see e.g. \cite{web11}. \\
In our measurement of muon lifetime we will use muons from cosmic rays. 
Cosmic rays have been introduced by Victor Hess in 1912 as an explanation of the observed 
increase of the environmental radiation with altitude \cite{hess}.
By that time, the existence of radioactive elements in the Earth had been established,
and they were considered as a source of natural background radiation. However, that could not explain 
the observed
increase of the radiation with altitude, and V. Hess proposed a cosmic origin of this radiation.
Subsequent investigations revealed the nature of this cosmic
radiation and led to the discovery of many new particles, such as positron - the first discovered
antiparticle, muon,
pion, strange particles (hadrons with strange quarks) and hypernuclei (nuclei that alongside ordinary 
nucleons contain also baryons with strange quarks).
Nowadays, the nature of cosmic rays is pretty well known, although there are still some
open questions, for example the nature of their high energy part is still being strongly
investigated, see e.g. \cite{he-cosmics}. Cosmic rays consist of high-energy particles that impinge on 
Earth's atmosphere from outer space, the so-called primary cosmic rays.
The charged primary cosmic rays consist mainly of protons 
(86\%), helium nuclei (11\%), nuclei of heavier elements up to uranium (1\%) and electrons (2\%),
with a small component of antiparticles which is produced through the interaction of the primary
cosmic rays with the interstellar matter, see e.g. \cite{perk}.

In the reactions with the nuclei in the atmosphere they generate secondary particles, such as pions. 
In the decays of charged pions, muons and neutrinos are produced:
\begin{eqnarray}
\pi^- \rightarrow \mu^- + \bar{\nu_{\mu}}   \nonumber \\
\pi^+ \rightarrow \mu^+ + \nu_{\mu}   \nonumber 
\end{eqnarray}
The production of muons occurs at approximately 10 km above Earth's surface. Taking into
account the speed limit given by the speed of light and 
knowing the muon lifetime, it is clear that without the effect of time dilation 
it would not be  possible for muons to reach Earth's surface.

 Average muon flux on Earth's surface is approximately 200 muons/m$^2$s 
and their average energy is about 2 GeV. 
There is a very small fraction
of muons with the energy
below approximately 30 MeV, that can be stopped in our scintillator, and which subsequently decay in it.

\section{The mean lifetime of muon and the principle of our measurement}

Decays of unstable nuclei and particles, including muons,
are characterized by the decay probability
 in a unit of time, $\lambda$. The mean lifetime of the particle, $\tau$, 
 is defined as $\tau=\frac{1}{\lambda}$. 
 Decays follow the exponential radioactive decay law:
\begin{equation}
 N(t)=N_0 e^{-\frac{t}{\tau}}
\label{eq:n}
\end{equation}
where $N_0$ is the number of unstable particles at the beginning, $t=0$, and $N(t)$ is the number of particles
that survive until time $t$.
We know that the mean lifetime of muon is  $\tau$ = 2.1969811 $\pm$ 0.0000022 $\mu$s \cite{pdg}. 

In the case of radioactive nuclei, if $e^{-\frac{t}{\tau}}$
changes significantly in a reasonable time,
 one can prepare a sample of certain number of nuclei and measure the activity, $A(t)$,
 which is the number of decays in a unit of time.
 The activity follows the same exponential time dependence as the number of nuclei:
\begin{equation}
 A(t)= \frac{dN(t)}{dt}=A_0 e^{-\frac{t}{\tau}},
\end{equation}
where $A_0$ is activity at time $t=0$.
From the obtained time dependence of the measured activity by fitting exponential function one can
determine the lifetime. 

In our case of  lifetime measurement of cosmic muons, we cannot enclose cosmic muons
 in a box and prepare a sample of certain initial number of muons. Instead, we
register muons that come randomly into our scintillator.
  Most of the muons pass through the scintillator, 
 but some of them stop and decay in it.
 Since $e^{-\frac{t}{\tau}}$ in the case of muons changes significantly over time, 
 we will use a quantity that corresponds to the activity
 for the determination of the muon lifetime.
 In our case, the entrance of a muon into the scintillator represents the time $t=0$. 
 It is a consequence of 
 the fact that the exponential radioactive decay law, which governs
the decay of muons, is independent of the measurement starting time.
Registering the time differences between the entrance of muons and their decay, that means
the appearance of the signals from the corresponding electrons (positrons) in the scintillator,
we can determine  the distribution of muons decaying 
at time $t$, $I(t)$. This corresponds to the activity, $A(t)$, in the case of
 decay of radioactive nuclei, and $I(t)$ follows the same exponential time dependence:
\begin{equation}
 I(t)=I_0 e^{-\frac{t}{\tau}}
  \label{eq:I}
\end{equation}
where $\tau$ is the muon lifetime and $I_0$ is the decay rate at $t=0$.
 By fitting exponential function
 to the measured distribution one can obtain the muon lifetime.
 
 Neutrino and antineutrino, which are also produced in the decay of a muon, leave the detector
 unobserved, because they interact with the weak force with the matter.

\section{Experimental setup and measurement}
\subsection{Scintillation detector}
For the detection of cosmic muons in our setup we are using plastic scintillator, produced by  
Amcrys, Ltd, Kharkov, Ukraine, type UPS-89. It is 
cylindrically shaped with the base diameter of 10 cm and the height of 30 cm.
All scintillator surfaces, except one cylinder base,
are wrapped in aluminum foil. The uncovered base is coupled to a 3-inch and 9-stage
photomultiplier tube, PMT, by using optical grease (we have used an older PMT, Philips XP5312/SN,
but any with the similar properties can be used).
The scintillator and the PMT 
are mounted within an aluminum  tube, which protects them from the room ambient light.

The PMT high voltage (HV) supply was custom built, and it provides a fixed high voltage of
-950 V. For it we designed printed circuit board (PCB) which uses CCFL inverter as a
primary voltage boost device. After CCFL inverter a two-stage Cockcroft-Walton multiplier is used. 
At the end of the chain are zenner diodes in order to limit and keep the high voltage well
defined and stable.
As the input, the HV supply requires 5 V and it is supplied by a USB cable either from our 
electronics, 
by a computer, or directly from the net by using an appropriate adapter.
The high voltage supply is placed in the box
that is mounted beneath the aluminum tube with the scintillator and the PMT.

\begin{figure}[h]
\resizebox{0.5\textwidth}{!} {\includegraphics{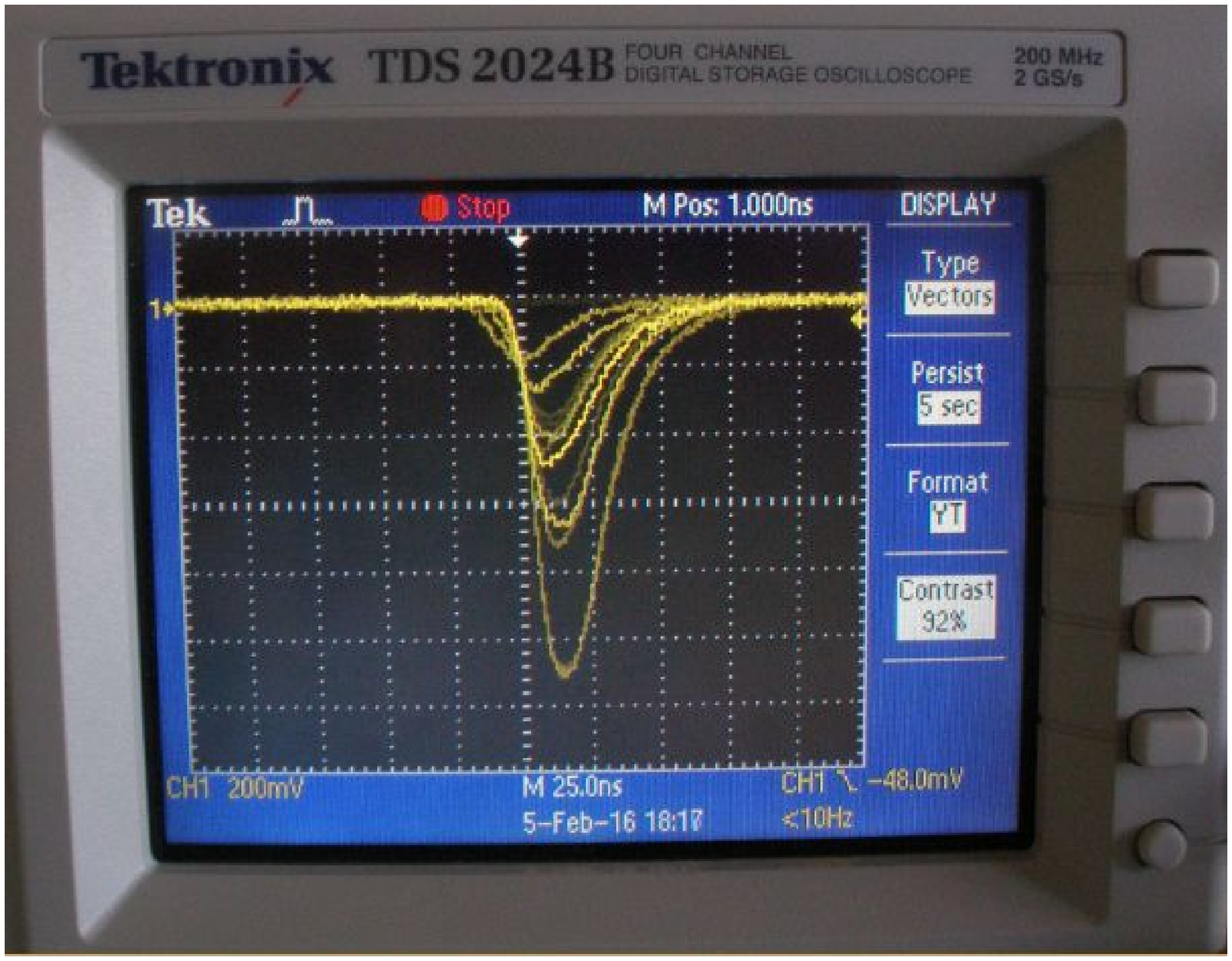}}
\resizebox{0.545\textwidth}{!} {\includegraphics{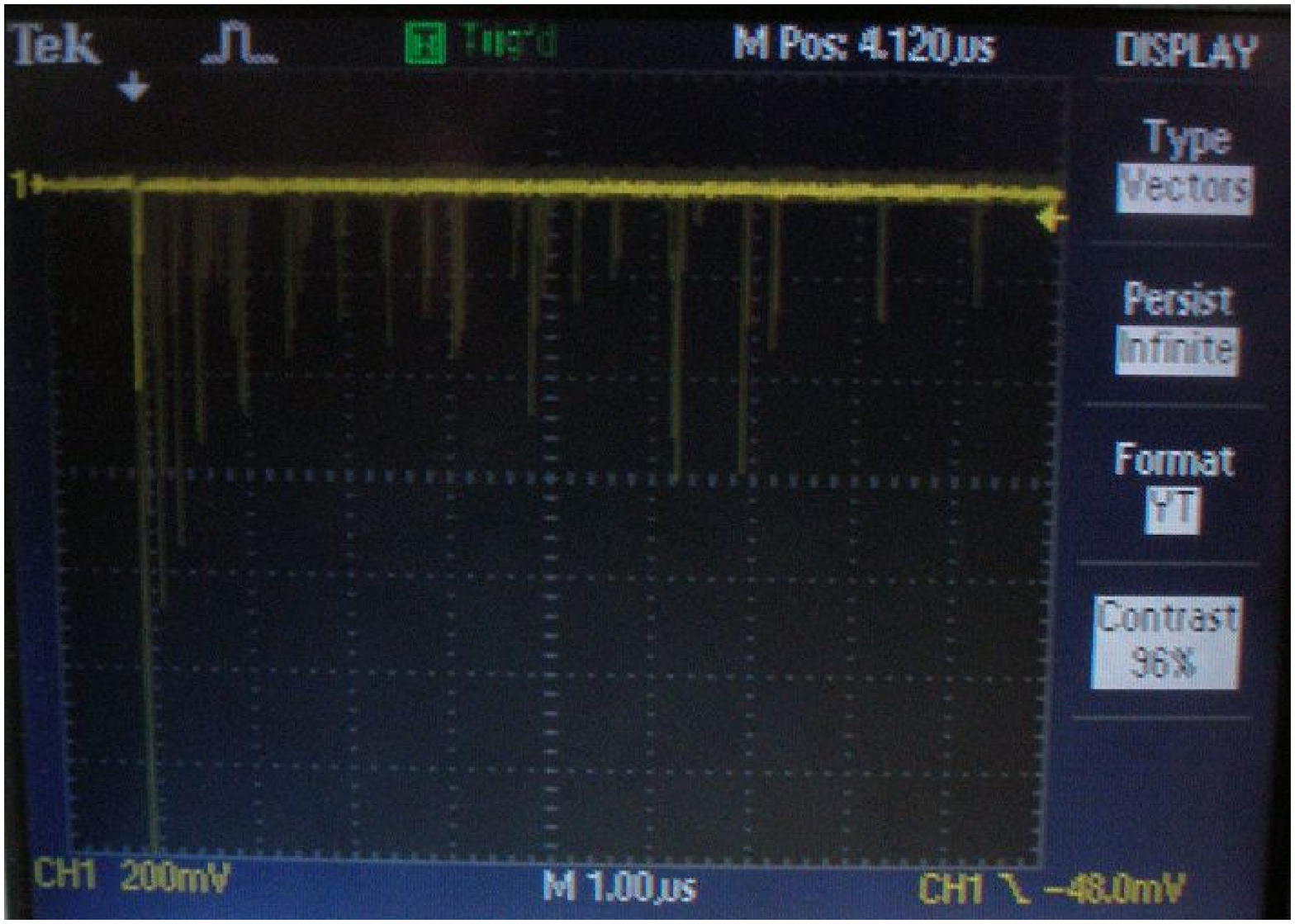}}
\caption{Muon signals from the scintillation detector observed on the oscilloscope with
the threshold of 48 mV (left). The time distribution of electrons from
 decays of muons (right). The trigger is given by an 
 incoming muon and the second signal is produced by the electron (or positron) from the decay of a stopped muon.}
\label{osci1}
\end{figure}

We monitored the negative signals from the scintillation detector,
which were mainly produced by cosmic muons, by a digital oscilloscope Tektronix TDS 2024B.
 The signals are shown in Figure~\ref{osci1}, left, with the oscilloscope working in the persistence
 mode. 
With the threshold of -48 mV, which eliminates the noise, the observed rate was several Hz,
as expected from the muon flux on the ground level and the dimensions of the scintillator.
These signals are produced mainly by cosmic muons, but among these events are also signals
produced by electrons from the decays of muons that are stopped in the scintillator and also 
background events. 
Using the time scale of the oscilloscope in the microsecond region, 
a decay of the stopped muon can be observed as the second signal on the oscilloscope
 caused by the electron from the decay in the scintillator.
Using this time scale  and the persistence mode, one can directly observe the
 time distribution of electrons from the
 decays of muons on the oscilloscope screen, Figure~\ref{osci1}, right.

 In principle, the simplest, but not very practical, way to determine the muon
 lifetime is to use this observed
 time distribution \cite{muehry}.
 Much more convenient way is to record this time distribution in a file for the further analysis. 
 
 \subsection{Readout electronics}
 Majority of previous setups has used conventional nuclear physics electronics, usually NIM 
 discriminators, to process signals from the scintillation detector and all have used
 some kind of TDC. In the TDC an incoming muon gives a start signal and the electron 
 from the decay gives a stop signal. We have followed a different path and have used a low cost
 electronics to record the time stamps of the events that are registered in the scintillation detector. 
 The muon lifetime can then be determined in the subsequent analysis of the recorded events. 
 
 In order to register the time of a signal appearance,
the signals from the scintillation detector are fed to a fast comparator.
For that purpose we designed a comparator PCB
module. We used an AD8561 fast TTL comparator, with a variable
threshold in the range of 0 to -500 mV, and after it the signal shape is modified by 
a 74121 chip, which gives constant width of the output pulse regardless of the input pulse width.
The output of the comparator module is a positive logical signal for each input signal above
the set threshold.
For pedagogical reasons, signals from the comparator module can also be monitored on the oscilloscope.
We set the threshold to -50 mV in order to discriminate the noise from the PMT and acquire 
 as much as possible of the physical signals.

 Signals from the comparator module are led to the custom designed timer PCB module
 which contains  32-bit PIC32 processor \cite{pic}.
 This module can record time stamps of the signals with a time resolution of 12.5 ns.
 The range of time measurement is approximately 54 seconds (2$^{32} \times$ 12.5 ns).
 After this time period, time is reset to zero and the measurement continues automatically until it is stopped
 by the user.
 The module can be programmed to transfer the recorded data to a computer or a memory card and
 also to perform simple operational tasks on the data.
In our case, once 16 events are 
recorded, the whole block of 16 time stamps is sent via serial/USB connection to the
computer.  

The PIC32 module was programmed to show the total number of registered events
during the measurement. It was also programmed to calculate the time differences between subsequent 
events and to show the number of events for which the time difference is less than
some selected time window, usually several muon lifetimes.
This number indicates the number of registered muon decays in the selected time window, as it will be
explained below. 
\begin{figure}[h]
\resizebox{0.5\textwidth}{!} {\includegraphics{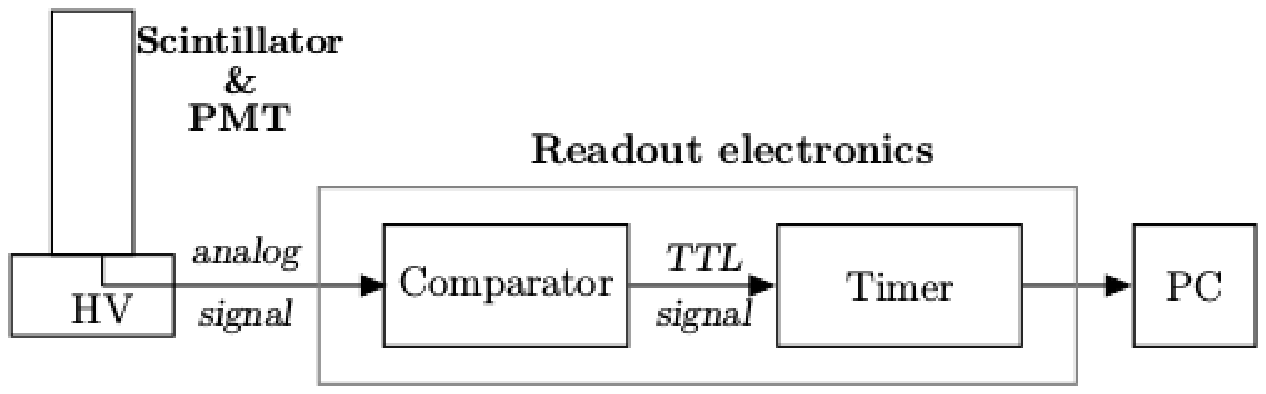}}
\resizebox{0.5\textwidth}{!} {\includegraphics{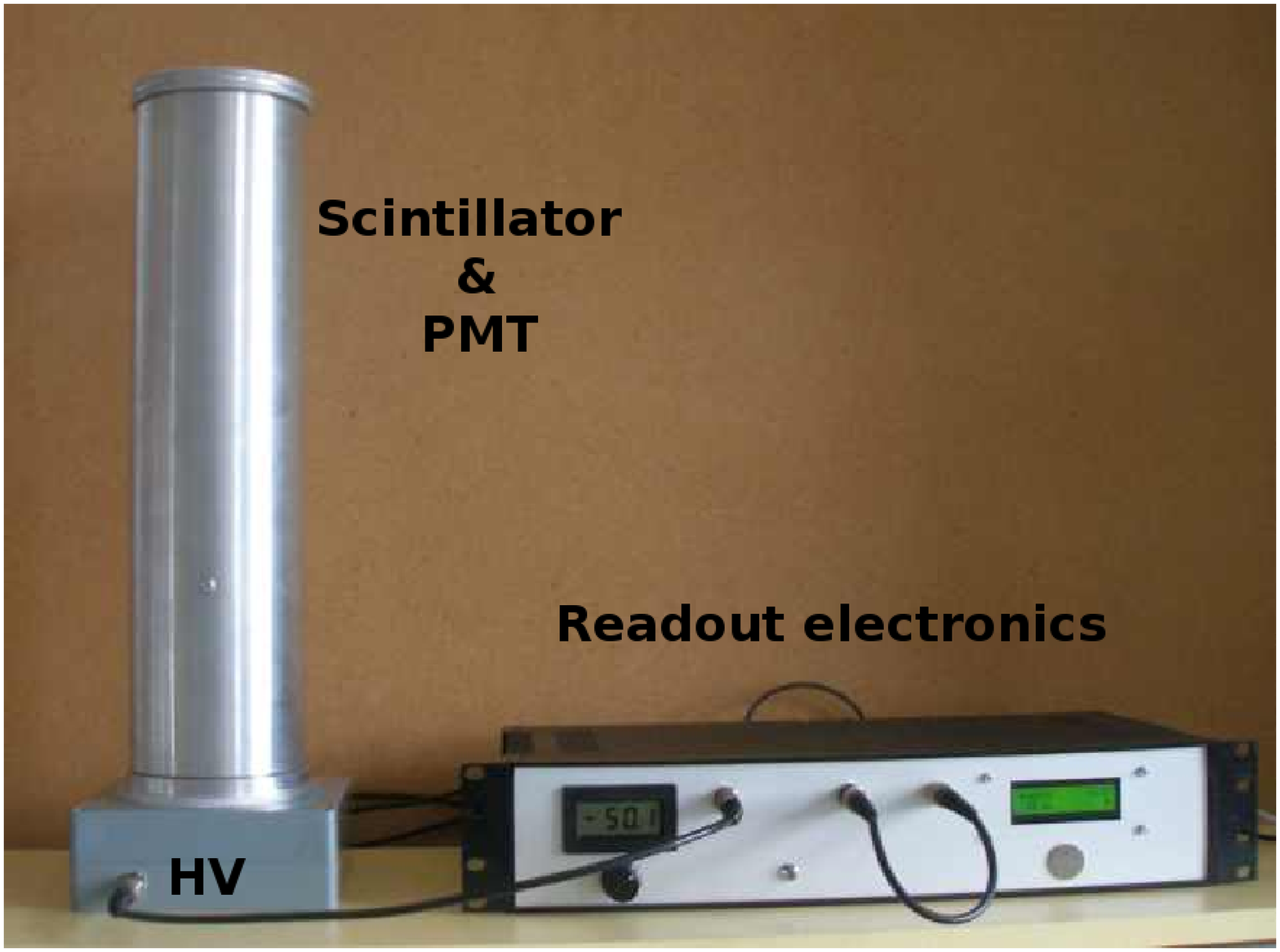}}
\caption{The scheme of the assembled setup (left). The scintillation
  detector and the readout electronics mounted in the box (right).}
\label{dspic}
\end{figure}

The scheme of the assembled setup is shown in Figure~\ref{dspic}, left, and
the scintillation detector and electronics mounted in the box are shown in Figure~\ref{dspic}, right.
The most expensive part in our setup is the scintillation detector, the standard
part for the detection of incoming muons and electrons from the decays 
in the majority of muon lifetime setups. However, the electronic components for signal processing and
data acquisition are more than an order of magnitude cheaper compared to the classical nuclear
physics equipment. Also, this electronics is much simpler and more compact. 


\section{Data analysis and results}
From the recorded time stamps of the registered events in the scintillation
detector we can determine the muon lifetime.
As the first step, we have to calculate 
the time differences between every pair of subsequently registered events.
Decays of the stopped muons can be selected
by constraining these time differences to be  smaller than some selected
time window, e.g. 10 $\mu$s. 
 On this way we obtain the time distribution of muon decays $I(t)$ defined in the equation (\ref{eq:I}).
 
In principle, we can determine the muon lifetime by fitting the exponential function to 
the obtained time distribution.
However, in the real measurement we have to take
final time intervals $\Delta T$ in which we count the number of muons that decay, $N_i$, in order to have 
a statistically significant number of counts.
It is desirable that the duration of the time interval $\Delta T$ is
less than $\frac{\tau}{4}$. It can be also shown that systematic uncertainty 
in determination of $\tau$ is less than 0.2~$\%$ if $\Delta T \le 0.1 \tau$ \cite{evans}.
On the other hand, if the time intervals are too small, this means a longer measurement
to collect sufficient statistics,  and a compromise should be found. 
The statistical uncertainty  of the number of muon decays in each time interval is calculated 
as $\sqrt{N_i}$. 

Finally,  we have to fit the exponential function
to the obtained data distribution to determine the muon lifetime $\tau$. 
The fit can be done either by using a simple exponential function (in the case we neglect 
a possible constant background) or by using a simple exponential in the combination with a
fixed term which takes into account the background \cite{mel}.

Students can program their own software package for the data analysis with different level of complexity 
or use some of the existing analysis tools.
We have developed a software package using a simple exponential function of the form given
by (\ref{eq:I}), and a graphical user interface (GUI) based
 on the Python software \cite{phyton} for the analysis of the collected data and the graphical 
representation of the results. 


\begin{figure}[h]
{\includegraphics[width=80mm]{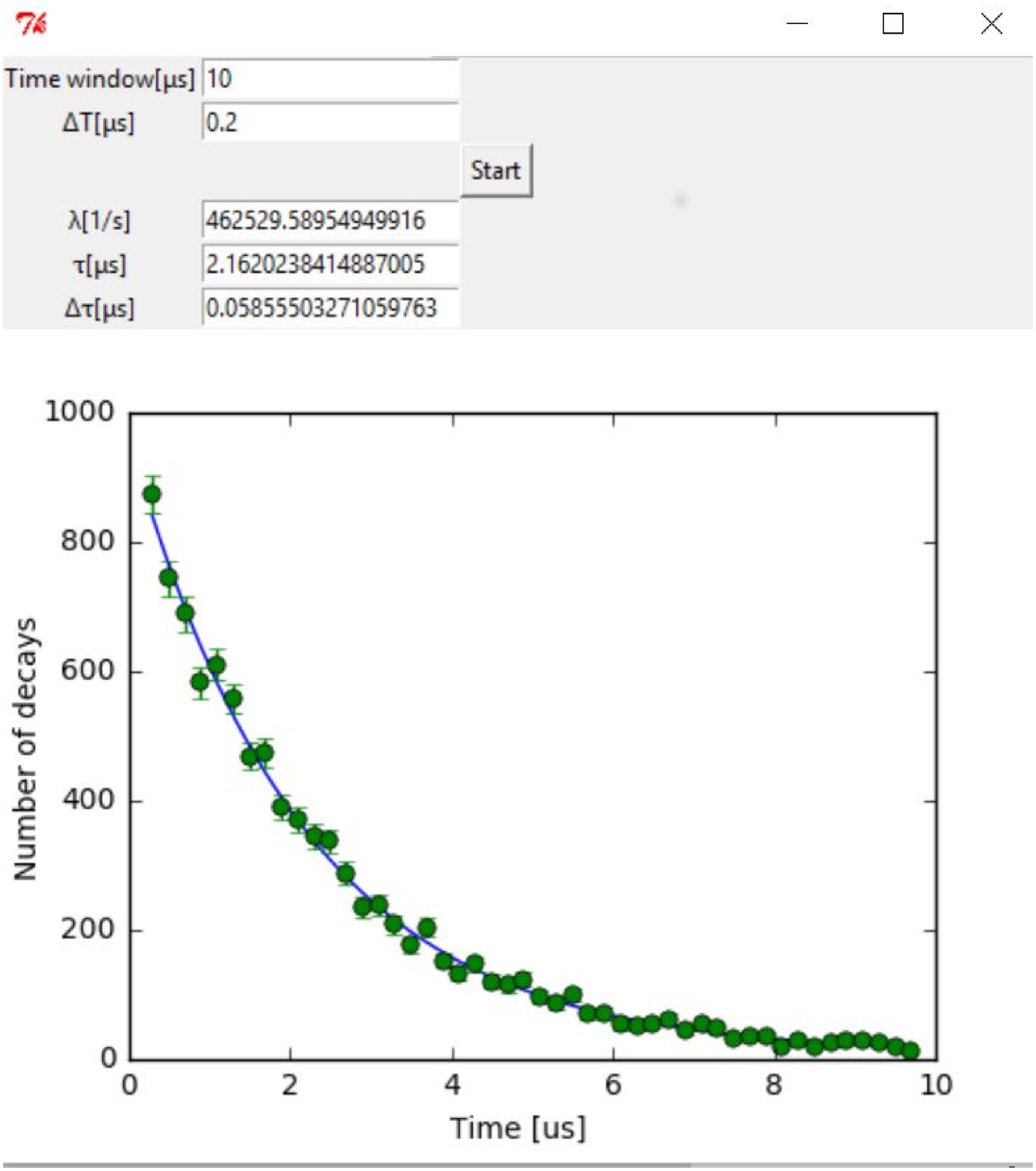}}
\caption{The graphical user interface for the data analysis and the representation of the results. 
In this case we took
the events that have time differences less than 10 $\mu$s and $\Delta$T was 0.2 $\mu$s. The numbers of muon 
decays in the given time interval, $N_i$, are
presented by green solid points and for each point was calculated the statistical uncertainty as $\sqrt{N_i}$.
The fitted curve is presented and the values of $\lambda$, $\tau$ and  uncertainty of $\tau$ are
given.}
\label{gui}
\end{figure}

During a single day of measurement with the setting described above, 
we collect approximately 1700 events 
in the time window of 10$\mu$s. The value of the time window in the analysis can be set by using the GUI.
To collect sufficient statistics, measurement needs to be performed during a course 
of several days.

In Figure~\ref{gui} we present the data collected in approximately 5 days of measurement, which
could be a reasonable time for a lab exercise. The time interval for counting the number of muon decays, 
$\Delta T$, which can also be set by using the GUI, is 0.2~$\mu$s 
(this is a little less than 0.1$\tau$).
The data distribution, the exponential fit on this data
distribution and the values of obtained $\lambda$, $\tau$ and uncertainty of $\tau$  are presented as the results
on the GUI screen.
The obtained result for the muon lifetime is $\tau=2.16\pm 0.06$ $\mu$s (statistical uncertainty only).

We have also seen that we can expect slight, but not much, improvements of the result with the increased 
statistics. On the other hand,
our setup has limitations caused by its very simple construction.
The signals are clean and above some (indeed very low) threshold there is no noise
from the scintillation detector and we have observed no noise from the electronics.
But, the biggest disadvantage is that we do not have particle identification, so both
muon and electron are detected by the same scintillation detector. However,
this is not so critical if we do not have any radioactive source in 
the vicinity of our detector, which means that we make a measurement in an ordinary environment.
In such an environment cosmic muons are the main source of radiation which is detected by the
detector. The condition that the second event appears in the chosen time window, in our case 10  $\mu s$,
after the muon entrance actually insures that we have detected the electron from the decay of
a muon stopped in the detector. Using the statistics of random events 
and knowing the rate of incoming cosmic rays as well as the dimensions of our scintillator,
one can show that there is very little probability that the second detected particle in the time
window of 10 $\mu s$ is cosmic ray particle, see e.g. \cite{mel}. We have estimated that there
is less than ten wrongly identified muon decays in a thousand of detected real decays. 
A possible background can be taken into account by using different fitting functions, such as
exponential plus constant term. 

One of the sources of the systematic uncertainties is also different
lifetime of negative muons in the matter compared to the vacuum, which is smaller because of
possible muon capture by nucleus, see  e.g. \cite{riggi}. 
Another possible source of systematic uncertainties is the finite size of the time window in which we 
seek the decays of muons. However, larger size of this time window means also bigger chance for
the second muon to appear in it and also at some point fluctuations in the background
become comparable with the number of  muon decays. The selected time window of 10 $\mu$s in our analysis
is slightly more than 4$\tau$, which is the time interval in which approximately 98.2\% of muons should decay.

\section{Conclusions}

We have assembled a simple and inexpensive setup for measurements of the lifetime of cosmic muons.
In our setup, as in the majority of other realizations of this experiment, we have retained the
scintillation detector for the detection of incoming muons and electrons from the decays, but we have
substituted classical nuclear physics equipment with inexpensive and simple electronics for signal processing
and data acquisition. This electronics is an order of magnitude cheaper, but still possesses required properties
for the measurement, and it is also simpler to use. This feature could enable much broader employment of
this experiment, not only in the undergraduate laboratory exercises, but also in 
secondary schools, although in some cases a help for the construction and operation would be needed.
All PCBs we have designed and programs we have developed, as well as additional information can be obtained on the request.

Another feature which differs our setup from, according to our knowledge, all other existing
setups is the principle on which the measurement is done.  We register time stamps of
all events which are detected in the scintillation detector and in the off-line analysis,
by putting appropriate cuts, the events which correspond to the decays of muons are 
selected. Since we have a kind of ``off-line TDC'', the time window in which we seek
the decay of a muon can be varied in the analysis.  

The software for the analysis can be developed either by the users themselves or they can use some
of existing tools. We have developed analysis software  and graphical user interface 
for the presentation of data and results by using Python software package. In our analysis software,
time intervals in which muon decays are counted can be varied as well as the time  window in which
we seek decays of muons. We have used only simple exponential function for 
the extraction of the muon lifetime. In the analysis of the  measurement of approximately five
days, with time interval of 0.2 $\mu$s and time window of 10 $\mu$s, we obtained a satisfactory 
result with the statistical uncertainty which is less than 3 \%. But, more advanced fitting procedures or
influence of the size of the time window in which we observe decays of muons or the size
of the time interval in which we count the decays can also be investigated.



\end{document}